# Conditional mode regression: Application to functional time series prediction

**Sophie Dabo-Niang[1] and Ali Laksaci[*2]**

[1]*Labo. GREMARS, Maison de Recherche,*
*Univ. Lille3,*
*BP60149, 59653 Villeneuve d'Ascq cedex Lille, France*
*e-mail:* `sophiedabo@univ-lille3.fr`

[2]*Département de Mathématiques*
*Univ. Djillali Liabès*
*BP 89, 22000 Sidi Bel Abbès, Algérie*
*e-mail:* `alilak@yahoo.fr`

**Abstract:** We consider $\alpha$-mixing observations and deal with the estimation of the conditional mode of a scalar response variable $Y$ given a random variable $X$ taking values in a semi-metric space. We provide a convergence rate in $L^p$ norm of the estimator. A useful and typical application to functional times series prediction is given.

## 1. Introduction

Let us introduce $n$ pairs of random variables $(X_i, Y_i)_{i=1,\ldots,n}$ that we suppose drawn from the pair $(X, Y)$, valued in $\mathcal{F} \times \mathbb{R}$, where $\mathcal{F}$ is a semi-metric space. Let $d$ denotes the semi-metric. Assume that there exists a regular version of the conditional probability of $Y$ given $X$, which is absolutely continuous with respect to Lebesgue measure on $\mathbb{R}$ and has bounded density. Assume that for a given $x$ there is some compact subset $S := (\alpha_x, \beta_x)$, such that the conditional density of $Y$ given $X = x$ has an unique mode $\theta(x)$ on $S$.

*corresponding author

In the remainder of the paper $x$ is fixed in $\mathcal{F}$ and $N_x$ denotes a neighborhood of $x$. Let $f^x$ (resp. $f^{x(j)}$) be the conditional density (resp. the $j^{th}$ order derivative of the conditional density) of the variable $Y$ given $X = x$. We define the kernel estimator $\widehat{f}^x$ of $f^x$ as follows:

$$\widehat{f}^x(y) = \frac{h_H^{-1} \sum_{i=1}^n K(h_K^{-1} d(x, X_i)) H(h_H^{-1}(y - Y_i))}{\sum_{i=1}^n K(h_K^{-1} d(x, X_i))}, \qquad \forall y \in \mathrm{I\!R}$$

with the convention $\frac{0}{0} = 0$. The functions $K$ and $H$ are kernels and $h_K = h_{K,n}$ (resp. $h_H = h_{H,n}$) is a sequence of positive real numbers. Note that a similar estimate was already introduced in the special case where $X$ is a real random variable by many authors, Rosenblatt (1969) and Youndjé (1996) among others. For the functional case, see Ferraty *et al.* (2006a). A natural and usual estimator of $\theta(x)$ denoted $\widehat{\theta}(x)$, is given by:

$$\widehat{\theta}(x) \;=\; \arg\sup_{y \in S} \widehat{f}^x(y). \tag{1}$$

Note that this estimate $\widehat{\theta}(x)$ is not necessarily unique, so, the remainder of the paper concerns any value $\widehat{\theta}(x)$ satisfying (1).

The main goal of this paper is to study the nonparametric estimate $\widehat{\theta}(x)$ of $\theta(x)$ when the explanatory variable $X$ is valued in the space $\mathcal{F}$ of eventually infinite dimension and when the observations $(Y_i, X_i)_{i \in \mathrm{I\!N}}$ are strongly mixing.

The motivation for this mode regression model is its interest in some nonparametric estimation problems where the mode regression provides better estimations than the classical mean regression (see for instance Collomb *et al.* 1987, Quintela & Vieu (1997), Ould-Sad (1997), Berlinet *et al.* (1998), or Louani & Ould-Sad (1999), for the multivariate case).

Currently, the progress of informatics tools permits the recovery of increasingly bulky data. These large data sets are available essentially by real time monitoring, and computers can manage such databases. The object of statistical study can then be curves (consecutive discrete recordings are aggregated and viewed as sampled values of a random curve) not numbers or vectors. Functional data analysis (FDA) (see Bosq 2000, Feratty and Vieu, 2006, Ramsay and Silverman, 1997, 2002) can help to analyze such high-dimensional data sets. The statistical problems involved in the modelization of functional random variables has received increasing interests in recent literature (see for example

Dabo-Niang 2002, Ferraty & Vieu 2004, Dabo-Niang & Rhomari 2004, Masry 2005 for nonparametric context). In this functional area, the first results concerning the conditional mode estimation were obtained by Ferraty *et al.* 2006a. They established the almost complete convergence of the kernel estimator in the i.i.d. case. This last result has been extended to dependent case by Ferraty *et al.* 2005. Ezzahrioui and Ould-said (2006a, 2006b) have studied the asymptotic normality of the kernel estimator of the conditional mode for both i.i.d. and strong mixing cases. The monograph of Ferraty & Vieu 2006b presents an important collection of statistical tools for nonparametric prediction of functional variables. Recently, Dabo-Niang & Laksaci (2007) stated the convergence in $L^p$ norm of the conditional mode function in the independent case.

In this paper, we consider the case where the data are both dependent and of functional nature. We prove the $p$-integrated consistency by giving the upper bounds for the estimation error. We show how our results can be applied to prediction of functional times series, by cutting the past of the time series in continuous paths. As an application, we applied our method to some environmental data.

The paper is organized as follows: the following Section is devoted to fixing notations and hypotheses. We state our results on Section 3. Section 4 is devoted to an application to a time series prediction problem.

## 2. Notation and Assumptions

We begin by recalling the definition of the strong mixing property. For this we introduce the following notations. Let $\mathcal{F}_i^k(Z)$ denote the $\sigma-$algebra generated by $\{Z_j, i \leq j \leq k\}$.

**Definition 1** *Let* $\{Z_i, i = 1, 2, ...\}$ *denote a sequence of rv's. Given a positive integer n, set*

$$\alpha(n) = \sup \left\{ |\mathbb{P}(A \cap B) - \mathbb{P}(A)\mathbb{P}(B)| : A \in \mathcal{F}_1^k(Z) \text{ and } B \in \mathcal{F}_{k+n}^\infty(Z), \ k \in \mathbb{N}^* \right\}.$$

*The sequence is said to be* $\alpha$*-mixing (strong mixing) if the mixing coefficient* $\alpha(n) \to 0$ *as* $n \to \infty$.

There exist many processes fulfilling the strong mixing property. We quote, here, the usual ARMA processes which are geometrically strongly mixing, *i.e.,* there exist $\rho \in (0,1)$ and $a > 0$ such that, for any $n \geq 1$, $\alpha(n) \leq a\rho^n$ (see, *e.g.*, Jones (1978)). The threshold models, the EXPAR models (see, Ozaki (1979)), the simple ARCH models (see Engle (1982)), their GARCH extension (see Bollerslev (1986)) and the bilinear Markovian models are geometrically strongly mixing under some general ergodicity conditions.

Throughout the paper, when no confusion is possible, we will denote by $C$ or $C'$ some strictly positive generic constants, and we will use the notation $B(x, h) = \{x' \in \mathcal{F} : d(x', x) < h\}$. Our nonparametric model will be quite general in the sense that we will just need the following assumptions:

(H1) $P(X \in B(x, r)) = \phi_x(r) > 0$.

(H2) $(X_i, Y_i)_{i \in \mathbb{N}}$ is an $\alpha$-mixing sequence whose coefficients satisfy

$$\exists a > 0, \exists c > 0 : \ \forall n \in \mathbb{N} \ \alpha(n) \leq cn^{-a}.$$

(H3) $\forall i \neq j$,

$$0 < \sup_{i \neq j} P\left[(X_i, X_j) \in B(x, h) \times B(x, h)\right] = O\left(\frac{(\phi_x(h))^{(a+1)/a}}{n^{1/a}}\right).$$

(H4) $\forall (y_1, y_2) \in S \times S$, $\forall (x_1, x_2) \in N_x \times N_x$,

$$|f^{x_1}(y_1) - f^{x_2}(y_2)| \leq C\left(d(x_1, x_2)^{b_1} + |y_1 - y_2|^{b_2}\right), \quad b_1 > 0,\ b_2 > 0.$$

(H5) $f^x$ is $j$-times continuously differentiable with respect $y$ on $S$ such that, $f^{x(l)}(\theta(x)) = 0$, for $1 \leq l < j$, and $\left|f^{x(j)}(y)\right| < \infty$, for all $y \in S$.

(H6) $K$ is a function with support $(0, 1)$ such that $0 < C' < K(t) < C < \infty$.

(H7) $H$ is a function which satisfies :
$$\begin{cases} \text{(i) There exists an integrable function } g \text{ such that} \\ \qquad |H(t) - H(s)| \leq Cg(|t - s|), \\ \text{(ii) } \int |t|^{b_2} H(t)dt < \infty \text{ , and } \quad \int H(t)dt = 1. \end{cases}$$

The concentration propriety (H1) is less restrictive that the fractal condition introduced by Gasser *et al.* (1998) and is known to hold for several continuous time processes (see for instance Bogachev (1999) for a gaussian measure, Li & Shao (2001) for a general gaussian process and Dabo-Niang & Laksaci (2007) for more discussion). In order to establish the same convergence rate as in the i.i.d. case (see Niang & Laksaci (2006) ), we reinforce the mixing by introducing (H2) and (H3). Note that we can establish the convergence results without these mixing assumptions, however, the convergence rate expression will be perturbed, it will contain the covariance term of the observations.

Assumptions (H4) and (H5) are regularity conditions which characterize the functional space of our model and are needed to evaluate the bias term in our asymptotic devlopments. It should be noted that, the flatness of the function $f^x$ around the mode $\theta(x)$ controlled by the number of vanishing derivatives at $\theta(x)$ ( assumption (H5)), has a great influence on the asymptotic rates of our estimate (see Theorem 1). Assumptions (H6) and (H7) are standard technical conditions in nonparametric estimation. They are imposed for the sake of simplicity and brevity of the proofs.

## 3. Main results

In this Section, we establish the $p$-mean rate of convergence of the estimate $\widehat{\theta}(x)$ to $\theta(x)$.

**Theorem 1** *Under the hypotheses (H1)-(H7), we have for all $p \in [j, \infty[$*

$$\left(E|\widehat{\theta}(x) - \theta(x)|^p\right)^{1/p} = O\left(h_K^{\frac{b_1}{j}} + h_H^{\frac{b_2}{j}}\right) + O\left(\left(\frac{1}{n\,\phi_x(h_K)}\right)^{\frac{1}{2j}}\right),$$

*whenever*

$$\exists \eta > 0, \qquad Cn^{\frac{2+p-a}{a+1-p}+\eta} \leq \phi_x(h_K) \leq C'n^{\frac{1}{1-a}} \qquad (2)$$

*holds with $a > \max\left(p+1, \left(4+p+\sqrt{(4+p)^2 - 4 - 8p}\right)/2\right)$.*

**Proof.** Let us now write the following Taylor expansion of the function $f^x$ under (H5)

$$f^x(\widehat{\theta}(x)) = f^x(\theta(x)) + \frac{1}{j!}f^{x(j)}(\theta^*)(\theta(x) - \widehat{\theta}(x))^j,$$

for some $\theta^*$ between $\theta(x)$ and $\widehat{\theta}(x)$. By simple analytic arguments, we can show that

$$|\widehat{\theta}(x) - \theta(x)|^j \leq \frac{j!}{\min\limits_{y \in (\alpha_x, \beta_x)} f^{x(j)}(y)} \sup\limits_{y \in (\alpha_x, \beta_x)} |\widehat{f}^x(y) - f^x(y)|.$$

The Minkowski inequality permits us to write:

$$\left\|\sup\limits_{y \in (\alpha_x, \beta_x)} \left|\widehat{f}^x(y) - f^x(y)\right|\right\|_p \leq \left\|\sum_{i=1}^n W_{ni}(x) \sup\limits_{y \in (\alpha_x, \beta_x)} |H_i(y) - E\left(H_i(y)/X_i\right)|\right\|_p$$

$$+ \left\|\sum_{i=1}^n W_{ni}(x) \sup\limits_{y \in (\alpha_x, \beta_x)} |E\left(H_i(y)|X_i\right) - f^x(y)|\right\|_p$$

$$+ \sup\limits_{y \in (\alpha_x, \beta_x)} |f^x(y)| \left(P\left(\sum_{i=1}^n W_{ni}(x) = 0\right)\right)^{1/p}, \ \forall p \geq j,$$

where

$$W_{ni}(x) = \frac{K(h_K^{-1}d(x, X_i))}{\sum_{i=1}^n K(h_K^{-1}d(x, X_i))} \quad \text{and} \quad H_i(y) = h_H^{-1}H(h_H^{-1}(y - Y_i)).$$

Then, Theorem 1 is a consequence of the following lemmas.

**Lemma 1** *We get under the hypotheses (H1), (H4), (H6) and (H7(ii)):*

$$\left\| \sum_{i=1}^n W_{ni}(x) \sup_{y \in \mathbb{R}} |E\left(H_i(y)| X_i\right) - f^x(y)| \right\|_p = O(h_H^{b_2}) + O(h_K^{b_1}).$$

**Lemma 2** *Under the hypotheses (H1)-(H3), (H6) and (H7(i)), we have for all $p \in [j, \infty[$*

$$\left\| \sum_{i=1}^n W_{ni}(x) \sup_{y \in \mathbb{R}} |H_i(y) - E\left(H_i(y)| X_i\right)| \right\|_p = O\left(\frac{1}{n\phi_x(h_K)}\right)^{\frac{1}{2}}$$

*whenever*

$$\exists \eta > 0, \qquad Cn^{\frac{2+p-a}{a+1-p}+\eta} \le \phi_x(h_K) \le C'n^{\frac{1}{1-a}} \tag{3}$$

*holds with $a > \max\left(p+1, \left(4+p+\sqrt{(4+p)^2-4-8p}\right)/2\right)$.*

**Lemma 3** *Under the conditions of Lemma 2, we have:*

$$\left(P\left(\sum_{i=1}^n W_{ni}(x) = 0\right)\right)^{1/p} = o\left(\frac{1}{n\phi_x(h_K)}\right)^{\frac{1}{2}}.$$

## 4. Application

The most important application of conditional mode estimation when the observations are dependent and of functional nature is the prediction of future values of some process by taking into account the whole past continuously. Indeed, let $(Z_t)_{t \in [0,b[}$ be a continuous time real valued random process. From $Z_t$ we may construct $N$ functional random variables $(X_i)_{i=1,\ldots,N}$ defined by:

$$\forall t \in [0, b[, \qquad X_i(t) = Z_{N^{-1}((i-1)b+t)},$$

and a real characteristic $Y_i = G(X_{i+1})$. The above consistency result permits to predict the characteristic $Y_N$ by the conditional mode estimate $\widehat{Y} = \widehat{\theta}(X_N)$ given by using the $N-1$ pairs of r.v $(X_i, Y_i)_{i=1,\ldots,N-1}$.

## 5. Appendix

**Proof of lemma 1** : By definition of the $L^p$ norm, we have

$$\left\| \sum_{i=1}^{n} W_{ni}(x) \sup_{y \in (\alpha_x, \beta_x)} |E(H_i(y)|X_i) - f^x(y)| \right\|_p$$

$$= E^{1/p} \left| \sum_{i=1}^{n} W_{ni}(x) \mathbb{1}_{B(x,h_K)}(X_i) \sup_{y \in (\alpha_x, \beta_x)} |E(H_i(y)|X_i) - f^x(y)| \right|^p,$$

where $\mathbb{1}$ is the indicator function. If we consider the change of variables $t = \dfrac{y-z}{h_H}$, then we get under (H7 (ii)) and (H4)

$$|E(H_i(y)|X_i) - f^x(y)| \leq \int \left| H(t) \left[ f^{X_i}(y - th_H) - f^x(y) \right] \right| dt,$$

$$\mathbb{1}_{B(x,h_K)}(X_i) |(E(H_i(y)|X_i) - f^x(y))| \leq C \left[ \int (|th_H|^{b_2} + |h_K|^{b_1}) H(t) dt \right].$$

We deduce from $\sum_{i=1}^{n} W_{ni}(x) = 1$, that:

$$\left\| \sum_{i=1}^{n} W_{ni}(x) \sup_{y \in (\alpha_x, \beta_x)} |E(H_i(y)|X_i) - f^x(y)| \right\|_p \leq C \left[ \int (|th_H|^{b_2} + |h_K|^{b_1}) H(t) dt \right].$$

The proves the lemma. ∎

**Proof of lemma 2**

It is easy to see that

$$\left\| \sum_{i=1}^{n} W_{ni}(x) \sup_{y \in \mathbb{R}} |H_i(y) - E(H_i(y)|X_i)| \right\|_p$$

$$\leq C E^{1/p} \left[ \left( \sup_j W_{nj} \right)^{p/2} \left( \sum_{i=1}^{n} W_{ni}^{1/2}(x) \sup_{y \in \mathbb{R}} |H_i(y) - E(H_i(y)|X_i)| \right)^p \right],$$

By (H7 (i)), the definition of $H_i$ and the boundedness of the conditional density with respect to the two variables, we have for all $i$:

$$|H_i(y) - E(H_i(y)|X_i)| = h_H^{-1} \left| H(h_H^{-1}(y - Y_i)) - \int H(h_H^{-1}(y - z)) f^{X_i}(z) dz \right|$$

$$\leq h_H^{-1} \left[ \int \left| H(h_H^{-1}(y - Y_i)) - H(h_H^{-1}(y - z)) \right| f^{X_i}(z) dz \right] \leq C \left[ \int h_H^{-1} g \left( |h_H^{-1}(Y_i - z)| \right) dz \right].$$

It follows from the usual change of variables $h_H^{-1}(Y_i - z) = t$ that

$$|H_i(y) - E(H_i(y)|X_i)| \leq C \int g(|t|) \, dt.$$

Moreover, by Cauchy-Schwartz inequality, we can write

$$\sum_{i=1}^n W_{ni}^{1/2}(x) \leq \sqrt{n}.$$

So, we deduce that

$$\left\| \sum_{i=1}^n W_{ni}(x) \sup_{y \in (\alpha_x, \beta_x)} |H_i(y) - E(H_i(y)|X_i)| \right\|_p \leq C\sqrt{n} \left[ \int g(|t|) \, dt \right] C' E^{1/p} \left( \sup_j W_{nj} \right)^{p/2}$$

We first evaluate the quantity $E^{1/p} \left( \sup_j W_{nj} \right)^{p/2}$. For this, set $U = K(h_K^{-1}d(X_n, x))$, $u = E(U)$ and $V = \sum_{j=1}^{n-1} K(h_K^{-1}d(X_j, x))$, it is clear that

$$C'\phi_x(h_K) \leq u \leq C\phi_x(h_K)$$

and $E(V) = (n-1)u$. If we consider the random variable $Z_{n-1} = \min(1, C/V)$, then, for all $j$

$$W_{nj}(x) = K(h_K^{-1}d(X_j, x))/(U + V) \leq Z_{n-1}.$$

We have for all $c > 0$;

$$Z_{n-1}^{p/2} = Z_{n-1}^{p/2} \mathbb{1}_{V < c} + Z_{n-1}^{p/2} \mathbb{1}_{V \geq c} \leq \mathbb{1}_{V < c} + (C/c)^{p/2}.$$

For $c = (n-1)u/2$, we get:

$$E(Z_{n-1}^{p/2}) \leq P(V < (n-1)u/2) + \left( \frac{C}{(n-1)u} \right)^{p/2}.$$

Clearly,

$$\begin{aligned}
P(V < (n-1)u/2) &= P(V - E(V) < -(n-1)u/2) \\
&\leq P(|V - E(V)| \geq (n-1)u/2) \\
&\leq P\left( \frac{1}{(n-1)EK_1} |V - E(V)| \geq 1/2 \right).
\end{aligned}$$

By using the Fuk-Nagaev's inequality (Rio 1999) we can show that, for all $\lambda > 0$ and $r > 1$

$$P\left( \frac{1}{(n-1)EK_1} |V - E(V)| \geq 4\lambda \right) \leq A_1 + A_2$$

where

$$A_1 = \left(1 + \frac{\lambda^2(n-1)^2(EK_1)^2}{rVar(V)}\right)^{-r/2} \text{ and } A_2 = 4\left(\frac{n-1}{r}\left(\frac{r}{\lambda(n-1)EK_1}\right)^{a+1}\right).$$

Set $\lambda = (\lambda_0/4)\sqrt{\dfrac{\log(n-1)}{(n-1)\phi_x(h_K)}}$, we get

$$A_2 \le C(n-1)r^a(n-1)^{-(a+1)/2}\phi_x(h_K)^{-(a+1)/2}(\log(n-1))^{-(a+1)/2}$$

By taking $r = O((\log(n-1))^2)$ and by using the left part of inequality (3), we can find $\eta' > 0$ such that

$$A_2 \le Cn^{-\eta'}n^{-p/2}\phi_x(h_K)^{-p/2}. \tag{4}$$

For $A_2$, we must evaluate asymptotically the quantity

$$Var(Z) = \sum_{i,j=1}^{n-1} Cov(K_i, K_j) := s_n^{2*} + nVar(K_1)$$

where

$$s_n^{2*} = \sum_{i \ne j}^{n-1} Cov(K_i, K_j).$$

In the sequel, we use techniques developed by Masry (1986) to give the asymptotic behavior of $s_n^{2*}$. Define the sets

$$S_1 = \{(i,j) \text{ such that } 1 \le i - j \le m_n\}$$

and

$$S_2 = \{(i,j) \text{ such that } m_n + 1 \le i - j \le n - 1\}$$

where the sequence $m_n$ is chosen such that $m_n \to \infty$. We denote by $J_{1,n}$ and $J_{2,n}$ be the sum of the covariance over $S_1$ and $S_2$ respectively. Then,

$$J_{1,n} = \sum_{S_1} |Cov(K_i, K_j)| \le \sum_{S_1} |EK_iK_j - EK_iEK_j|.$$

Because of (H1), (H3) and (H6) we can write

$$J_{1,n} \le Cnm_n\phi_x(h_K)\left(\left(\frac{\phi_x(h_K)}{n}\right)^{1/a} + \phi_x(h_K)\right).$$

On the other hand, to study the sum over $S_2$, we use the Davydov-Rio's inequality (see Rio, (1999)) in the $L^\infty$ cases. This leads, for all $i \ne j$ to

$$|Cov(K_i, K_j)| \le C\alpha(|i-j|)$$

and therefore we get

$$J_{2,n} = \sum_{S_2} |Cov(K_i, K_j)| \le C n^2 m_n^{-a}.$$

By choosing $m_n = \left( \dfrac{\phi_x(h_K)}{n} \right)^{-1/a}$, using the right part of (3), we obtain

$$\sum_{i \ne j}^{n-1} Cov(K_i, K_j) = O(n\phi_x(h_K)). \tag{5}$$

Now, for all $i = 1, \dots n - 1$ we can write

$$Var(K_1) = E(K_1^2) - (EK_1))^2.$$

By using (H1) we get

$$Var(K_1) \le C(\phi_x(h_K) + (\phi_x(h_K))^2).$$

Finally, this last result combined with (5) leads directly to

$$Var(Z) = O((n-1)\phi_x(h_K)). \tag{6}$$

This allows us to deduce that

$$A_1 \le C \left( 1 + \frac{\lambda_0^2 \log(n-1)}{16r} \right)^{-r/2} = C \exp\left( -r/2 \log\left( 1 + \frac{\lambda_0^2 \log(n-1)}{16r} \right) \right)$$

Choosing $r = C(\log(n-1))^2$, yields that

$$A_1 \le C \exp\left( -\lambda_0^2 \frac{\log n}{32} \right) = C n^{-\lambda_0^2/32}.$$

As $(p < \infty)$, for $\lambda_0$ large enough,

$$A_1 \le C n^{-\lambda_0^2/32} \le C n^{-\nu} n^{-p/2} \phi_x(h_K)^{-p/2} \text{ for some } \nu > 0. \tag{7}$$

So, by combining the results (4) and (7) we get, for $\lambda_0$ large enough,

$$P(V < (n-1)u/2) = O\left( \left( \frac{1}{n\phi_x(h_K)} \right)^{p/2} \right) \tag{8}$$

which implies that

$$E^{1/p}(Z_{n-1}^{p/2}) = O\left( \left( \frac{1}{n\phi_x(h_K)} \right)^{1/2} \right).$$

So,

$$\left\| \sum_{i=1}^n W_{ni}(x) \sup_{y \in \mathbb{R}} |H_i(y) - E(H_i(y)|X_i)| \right\|_p = O\left( \left( \frac{1}{n\phi_x(h_K)} \right)^{1/2} \right).$$

∎

**Proof of lemma 3**

We set $Z = \dfrac{1}{EK_1} \sum\limits_{i=1}^{n} K_i$, then, $\left( P \left( \sum\limits_{i=1}^{n} W_{ni}(x) = 0 \right) \right) = (P(Z = 0))$. It is clear that, for all $\varepsilon < 1$, we have

$$
\begin{aligned}
(P(Z = 0)) \quad & \leq (P(Z \leq 1 - \epsilon)) \\
& \leq (P(|Z - 1| \geq \epsilon)).
\end{aligned}
$$

By following similar arguments to those involved in the proof of (8), we have that:

$$
P(|Z - 1| \geq \epsilon) = o \left( \left( \frac{1}{n\phi_x(h_K)} \right)^{p/2} \right).
$$

This implies that,

$$
\left( P \left( \sum_{i=1}^{n} W_{ni}(x) = 0 \right) \right)^{(1/p)} = o \left( \left( \frac{1}{n\phi_x(h_K)} \right)^{1/2} \right).
$$

∎

*Acknowledgements :* The authors thank Professors F. Ferraty and Ph. Vieu of the LSP, University of Paul Sabatier, and N. Rhomari of University of Oujda, for helpful discussions and comments.